\documentstyle[12pt]{article}
\evensidemargin \oddsidemargin
\def\gappeq{\mathrel{\rlap {\raise.5ex\hbox{$>$}}
{\lower.5ex\hbox{$\sim$}}}}
 
\def\lappeq{\mathrel{\rlap{\raise.5ex\hbox{$<$}}
{\lower.5ex\hbox{$\sim$}}}}
 
\begin{document}
 
\begin{titlepage}
\begin{flushright}
OUTP-96-47P \\
hep-th/9607192
\end{flushright}

\begin{centering}
\vspace{.1in}
{\large {\bf Non-trivial Infrared 
Structure in (2+1)-dimensional Quantum Electrodynamics 
}} \\
\vspace{.2in}
{\bf I.J.R. Aitchison }, {\bf G. Amelino-Camelia},
{\bf M. Klein-Kreisler}$^\dagger$, 
{\bf N.E. Mavromatos}$^{*}$ 
and {\bf D. Mc Neill}
\\
\vspace{.1in}
\vspace{.03in}
Department of Physics
(Theoretical Physics), University of Oxford, 1 Keble Road,
Oxford OX1 3NP, U.K.  \\
\vspace{.1in}
{\bf Abstract} \\
\vspace{.05in}
\end{centering}
{We show that the gauge-fermion interaction
in multiflavour $(2 + 1)$-dimensional
quantum electrodynamics with a finite 
infrared cut-off
is responsible for non-fermi liquid behaviour in the infrared,
in the sense of leading to the existence of a non-trivial
fixed point at zero momentum, 
as well as to a significant slowing down 
of the running of the coupling at intermediate
scales as compared with previous 
analyses on the subject.
Both these features constitute deviations
from fermi-liquid theory. 
Our discussion is based on the leading- $1/N$ resummed
solution 
for the wave-function 
renormalization 
of the
Schwinger-Dyson equations .
The present work completes and confirms the expectations of an 
earlier work by two of the authors
(I.J.R.A. and N.E.M.) on the non-trivial infrared
structure of the theory.}
\vspace{0.2in}
%
\paragraph{}

\vspace{0.1in}
\begin{flushleft}
$^\dagger$On leave from Instituto de Fisica, UNAM, 
Abdo. Postal 20-364, 01000 Mexico DF, Mexico.
$^{*}$P.P.A.R.C. Advanced Fellow \\
July 1996 \\
\end{flushleft}
\end{titlepage}
\newpage
\paragraph{}
Recently two of us~\cite{am}, in a paper from now on to be 
referred as I, have suggested that 
a Schwinger-Dyson (SD) analysis for the self-energy 
of the massless fermion of $QED_3$ provided intriguing evidence 
for a non-trivial 
infrared structure
of the theory.We argued that the effective dimensionless 
running coupling constant varied slowly over a wide range 
of intermediate momenta lying between $\alpha$ (the 
mass scale of $QED_3$,which plays the r\^ole of an
ultra-violet cut-off
~\cite{app}) and the infrared cut-off $\epsilon$. 
This situation resembles that in (four-dimensional) 
`walking technicolour' models of particle physics,and we 
described it in terms of a `quasi-fixed point'.
The
inclusion of wave-function renormalization
was crucial to the effect. Equally important
was the incorporation of a finite infrared cut-off.
\paragraph{}
This situation is very interesting from 
a condensed-matter point of view. 
There have been arguments~\cite{dm} 
that a variant of $QED_3$ may serve as a continuum model
for simulating the  
physics of the novel high-temperature superconductors.
On the one hand, 
dynamical mass generation provides a mechanism 
for 
the spontaneous breaking of electromagnetic $U(1)$
symmetry, thus 
leading to superconductivity.
On the other , deviations from trivial infrared fixed point structure 
have recently been argued to be responsible for departures
from non-fermi liquid behaviour in the normal phase~\cite{shankar,am}
in which no dynamical mass is generated. 
Indeed 
the presence of wave-function 
renormalization 
yields marginal $O(1/N)$-corrections
to the `bulk' non-fermi liquid behaviour
caused by the gauge interaction in the limit of
infinite flavour number $N \rightarrow \infty$. 
Such corrections
lead to the appearance of modified critical
exponents.
In particular, at low temperatures there appear to be
logarithmic scaling violations of the linear resistivity
of the system of order $O(1/N)$~\cite{am}.
\paragraph{}
In I we based our arguments on an approximate form 
of the leading-$1/N$ resummed 
SD equations, which was an improved 
version of the results of ref. \cite{kondo}.
The approximation employed in \cite{kondo}
was strictly valid for a regime of momenta
$p << \alpha $,while that in \cite{am} was 
appropriate to $p \simeq \alpha$.
For the actual physical models, however, 
the regime of momenta over which deviations
from fermi liquid behaviour is observed 
is much wider than those employed in ref. \cite{kondo}
or ref.\cite{am}. 
To extend the validity of the arguments of I 
over a 
wider momentum regime in a mathematically 
rigorous way
one needs to 
have a more precise picture of the situation 
by employing as exact expressions for the $O(1/N)$ SD equations
as possible. In particular we would like to 
re-examine the way the effective coupling constant 
runs in the momentum range $0 < p < \alpha$, in 
a regime of parameters such that there is no 
dynamical mass generation. 
That is the purpose of this note. As we shall see,
the situation is exactly as predicted by the 
preliminary approximate analysis of I, thereby
confirming the expectations of that reference. 
\paragraph{}
To this end, it seems instructive first to recapitulate
the arguments of I while briefly reviewing the formalism. 
The SD equations, in the one-loop resummed $1/N$ limit,
read 
\begin{equation}
A(p)=1 - \frac{\alpha}{\pi^2N}\frac{1}{p^3} \int 
_0^\infty dk \frac{kA(k)G(p^2,k^2)}{k^2A(k)^2 + B(k)^2}I(p,k),
\label{one}
\end{equation}
where 
\begin{eqnarray}
&~&I(p,k) \equiv 
 \alpha ^2 ln\frac{p+k+\alpha}{|p-k|+\alpha} - 
\alpha (p+k - |p-k|)+ 2pk - \nonumber \\
&~&\frac{1}{\alpha}|p^2-k^2|
(p+k-|p-k|)-
\frac{1}{\alpha ^2}(p^2-k^2)^2\{ln\frac{p+k+\alpha}{|p-k|+\alpha}
-ln\frac{p+k}{|p-k|}\},  
\label{kernel}
\end{eqnarray} 
and 
\begin{equation}
B(p)=\frac{\alpha}{\pi^2 N} \frac{1}{p}\int_0^\infty dk
\frac{kB(k)G(p^2,k^2)}{k^2A(k)^2 + B(k)^2}\{4ln\frac{p+k+\alpha}
{|p-k|+\alpha}\} 
\label{oneb}
\end{equation}
where the Landau gauge has been assumed, and 
$\alpha \equiv \frac{e^2 N}{8}$ is the dynamically-generated
scale of the theory. 
$A(p)$ is the wave-function renormalization,
$B(p)$ is the mass-gap function, 
and $G(p^2,k^2)$ is a vertex function.  
The integrals in (\ref{one}) and (\ref{oneb}) 
are effectively cut-off
at $\alpha$, due to a sharp decay of the integrands
above this scale~\cite{app}. Hence $\alpha$ may be 
considered as an effective UV cut-off.
Moreover, for reasons that will become 
clear below we also introduce an infrared cut-off
$\epsilon$.
Following ref. \cite{kondo}, we make
the vertex ansatz
\begin{equation}
\Gamma _\mu (p^2,k^2) = \gamma _\mu A(k)^n \equiv \gamma _\mu G(k^2).
\label{three}
\end{equation}
The Pennington and Webb \cite{pen} ansatz corresponds to
$n=1$, where chiral symmetry breaking occurs for
arbitrarily large $N$ \cite{pis}. It is this case that
was argued to be consistent with the
Ward identities that follow from gauge invariance \cite{pen}.
Here we shall concentrate on the
$n=1$ ansatz.
It is interesting to note from (\ref{one}),(\ref{oneb}) 
that in this
case the right hand sides of the SD equations depend
only on the ratio $B(p)/A(p) \equiv \Sigma (p)$, which 
is the physical gap function.
\paragraph{}
For the purposes of I, and of this note, we are interested
in the normal phase of the theory
where the mass gap $B(p)$, or equivalently $\Sigma (p)$,
can be ignored in  the 
denominators of (\ref{one}). 
In that case,equation (\ref{one}) for $A(p)$ 
reduces to simple quadrature if $n=1$.
Using the ansatz (\ref{three}), one
can then analyze the Schwinger-Dyson (SD) equations,
in the various regimes of momenta,
in terms of a running
coupling obtained from 
substituting the solution for $A(p)$ 
into  equation (\ref{oneb})  for the gap:
\begin{equation}
g_R(p, \epsilon ) = \frac{g_0}{A(p, \epsilon )}
\label{renorm}
\end{equation}
where $g_0 =8/\pi ^2 N$, 
$N$ is the number of fermion
flavours, and $\epsilon$ is an infrared cutoff.
The reader might have wondered whether 
the definition (\ref{renorm}) of a running 
coupling makes any sense in the normal phase 
of the model, where dynamical mass generation 
is absent.The resolution of this comes from treating
the gap equation (\ref{oneb}) as a non-trivial equation,
of which the vanishing gap appears as a consistent 
solution.This provides a definition of $g_R$ which
continues smoothly from the phase of dynamical mass 
generation to the normal phase.
 Furthermore, the definition 
(\ref{renorm}) of the running 
coupling is also justified 
within the conventional 
framework of Gell-Mann-Low 
renormalization of the gauge model. 
\paragraph{}
The following two approximations 
were suggested in ref. \cite{kondo}:
\begin{eqnarray} 
&~&ln\frac{p+k+\alpha}{|p-k|+\alpha}
\simeq 2\{ \frac{k}{p+\alpha}+ \frac{1}{3}
(\frac{k}{p+\alpha})^3+ \dots \}\Theta (p-k)
+ \nonumber \\
&~&2 \{\frac{p}{k+\alpha}+ \frac{1}{3}
(\frac{p}{k+\alpha})^3 + \dots \}\Theta (k-p) 
\label{10a}
\end{eqnarray}
and 
\begin{eqnarray} 
&~&ln\frac{p+k+\alpha}{|p-k|+\alpha}\simeq
2 \{ \frac{k}{\alpha}-\frac{kp}{\alpha^2}
+\frac{3pk^2+k^3}{3\alpha^3}\}\Theta (p-k) +\nonumber \\
&~&2\{\frac{p}{\alpha}-\frac{pk}{\alpha^2}
+\frac{3pk^2+p^3}{3\alpha^3}\}\Theta (k-p).
\label{10b}
\end{eqnarray}
Using  (\ref{10b}),and neglecting $B(k)$ in the 
denominator of (\ref{one}),
the following 
approximate form for $A(p)$
in (\ref{one}) may then be derived~\cite{kondo}
\begin{equation}
A(p) = 1 - \frac{g_0}{3} \int _\epsilon ^\alpha dk
\frac{G(k^2) }{k A (k)}
\{(\frac{k}{p})^3 \Theta (p - k) + \Theta (k - p) \}. 
\label{sdir}
\end{equation}
Note that the form of the infrared regulator, which has now been 
introduced explicitly, is--following
 ~\cite{kondo}--simply a momentum-space 
cut-off. 
In \cite{kondo} an approximate expression 
for $g_R$ was given which was 
based on (\ref{sdir}), but with the further approximation 
of retaining only the term with $\Theta (k-p)$, while 
neglecting 
the one involving $\Theta (p-k)$. This produces the 
result (always for $n=1$ in (\ref{three}))
\begin{eqnarray}
g_R^{KN} &=&\frac{g_0}{1 + \frac{g_0}{3}ln(p/\alpha)}
\qquad {\rm for}~\epsilon < p < \alpha \nonumber \\
~&=&\frac{g_0}{1 + \frac{g_0}{3}ln(\epsilon/\alpha)} 
\qquad {\rm for}~0 < p < \epsilon. 
\label{grkn}
\end{eqnarray}
In I, by contrast, the term $\Theta (p-k)$ in 
(\ref{sdir}) was retained and the one 
with $\Theta (k-p)$ was dropped, leading 
to 
\begin{equation}
g_R^{AM}=\frac{g_0}{1-\frac{g_0}{9}+\frac{g_0}{9}(\epsilon/p)^3},
\label{gram}
\end{equation}
an approximation which was argued to be valid for $p \simeq \alpha$. 
Whereas $g_R^{KN}$ grows as $p$ decreases,
$g_R^{AM}$ exhibits the opposite behaviour, 
and this was interpreted in I as indicating 
that the true dependence of $g_R$ on $p/\alpha $
would be significantly slower than that of $g_R^{KN}$ -
a situation we described in terms of a `quasi fixed 
point structure '.
\paragraph{}

It is clearly
important  to establish the extent to which the 
conclusions drawn from (\ref{grkn}) or (\ref{gram}) 
are in fact reliable. Of course, a numerical 
solution of (\ref{one}) and (\ref{oneb}) 
is always possible (see below), but more 
insight can often be gained from analytical 
results, provided they are not artefacts
of a too-crude approximation scheme. 
In both I and \cite{kondo} the authors 
were interested in exploring the effects
of different $n$'s in (\ref{three}). 
But, as stated above, the case $n=1$
seems the most relevant. 
The main purpose of the present note is
to exploit the simplifications associated 
with the $n=1$ case to investigate more 
rigorously the issues raised in I and \cite{kondo}.
\paragraph{}
Before embarking on the calculations,we remark that for simplicity
we shall present results only for one value of $N$ (or $g_0$).
We select the number of flavours $N$ to lie comfortably
in the regime where dynamical mass is not generated.
 From the analysis of ~\cite{kondo} and~\cite{aklein} 
it becomes clear
that there is a critical number of flavours for dynamical 
mass generation to occur, and that this depends on the 
infrared cut-off. Although at present we have not 
studied the corresponding critical line in any detail,
we can be sure that the value $N=5$ will exceed the critical 
line for the regime of infrared cut-offs that are 
physically interesting to us ~\cite{am}, i.e.
$10^{-3} < \delta / \alpha < 10^{-1}$.We shall
present results for the value $N=5$.

\paragraph{} The simplest improvement of I and \cite{kondo} is 
obviously to retain both $\Theta $ terms 
in (\ref{sdir}). We easily find (for $n=1$)
\begin{equation}
g_R^{(\ref{sdir})}=g_0/A^{(\ref{sdir})}(p,\epsilon )
\label{gr5}
\end{equation}
with 
\begin{eqnarray}
A^{(\ref{sdir})} &=& 1-\frac{g_0}{9} +
\frac{g_0}{9}(\frac{\epsilon}{p})^3 + 
\frac{g_0}{3}ln(\frac{p}{\alpha}) \qquad {\rm for }~\epsilon < p <
\alpha \nonumber \\
&=& 1 + \frac{g_0}{3}ln(\frac{\epsilon}{\alpha}) \qquad {\rm for}~0 <
p < \epsilon 
\label{a5}
\end{eqnarray}
which is simply the obvious combination
of (\ref{grkn}) and (\ref{gram}). 
The quantities $g_R^{(\ref{sdir})}$ and 
$g_R^{KN}$ are shown 
in figure 1, from which we see that, as compared with 
(\ref{grkn}), (\ref{a5}) provides essentially a smooth approach 
to the region $0 \le p \le \epsilon $ where $g_R$ is constant 
(in this approximation). In this region, there is a line 
of non-trivial infrared fixed points, corresponding to the 
value $g^*_R = g_0/(1 + \frac{g_0}{3}ln(\epsilon/\alpha))$.
By demanding positivity of $A^{(\ref{sdir})}(p,\epsilon )$ at $p=\epsilon$
one can obtain a mild restriction on $\epsilon $:
\begin{equation}
\epsilon \ge 
e^{-\frac{3}{g_0}}\alpha
\label{bound}
\end{equation}
which is reasonable for $g_0 << 1$. 
\paragraph{}
The existence of a non-trivial infrared fixed point 
was one of the main conjectures of I, relevant 
for the non-fermi liquid behaviour of the system. 
Actually the same structure is evident in the $n=1$
curve of figure 1 of \cite{kondo}, 
though it was not remarked upon there; a similar 
conjecture was also made previously for four-dimensional 
$QCD$ in ref. \cite{higashijima}. 
The second conjecture made in I, also implying a non-fermi liquid behaviour,
pertains to a significant
slowing down of the rate of decrease of $g_R$ with increasing $p$
at intermediate scales of momenta,
relative to that shown by $g_R^{KN}$. 
Such behaviour is not apparent from figure 1; indeed,
$g_R^{(\ref{sdir})}$and $g_R^{KN}$ run essentially
parallel until $p$ is quite close to $\epsilon$.  
However, the conjectured behaviour at intermediate 
values of $p$ will in fact emerge from the more accurate
treatment which now follows. As we shall see, the 
restriction (\ref{bound}) also disappears, and one gets a 
non-trivial infrared fixed point structure at $p=0$. 
\paragraph{}
In order to improve upon (\ref{gr5}),
we return to (\ref{one}) with $B=0$, $n=1$, and a momentum-space
cut-off $\epsilon$. The running of the coupling $g_R (p,\epsilon)$
is then defined by (\ref{renorm})
with 
\begin{equation} 
A(p, \epsilon)= 1 -\frac{\alpha}{\pi^2 N}\frac{1}{p^3}
\int _\epsilon ^\infty \frac{dk}{k} I(p,k)
\label{improvedeps}
 \end{equation}
which we have evaluated numerically. The resulting
$g_R(p,\epsilon)$ is also shown in figure 1,
for comparison with $g_R^{(\ref{sdir})}$ and $g_R^{KN}$. 
It is indeed apparent that the exactly 
evaluated $g_R(p,\epsilon)$ does run  
significantly more slowly than either of  the approximate 
quantities $g_R^{(\ref{sdir})}$ and $g_R^{KN}$, thus confirming 
the second conjecture of I. 
\paragraph{}
In the discussion thus far, we have implemented the 
infrared regularization by a simple momentum-space cut-off
$\epsilon$. It may be questioned whether it is consistent,
in this case,to consider values of the external momentum $p$
lying below $\epsilon$,as we have done in figure 1. Indeed,
the behaviour of $g_R(p,\epsilon)$ for $p$ near $\epsilon$
and below it is,in our view,unlikely to be physical. Detailed 
examination shows that  
 $dg_R/dp$ vanishes at $p^*= \epsilon
(1 + {\cal O}[\epsilon/\alpha])$,
where $g_R$ reaches a maximum. However, for smaller values 
of the external momenta there is 
a decrease of  $g_R$ 
towards $p=\epsilon$, so that $p^*$ is not 
a fixed point. For values of $p$ less than $\epsilon$,
$g_R$ decreases further,leading ultimately (it appears)
to a fixed point at $p=0$. This behaviour seems unphysical,
and may signal an inconsistency between the vertex 
ansatz of (\ref{three}) and the momentum-space 
cut-off scheme, at small momenta. We would like to 
find an infrared regularization in which both $p$ and the 
integration variable $k$ could range between $0$ and $\alpha$,
and which would be expected to lead to smoother results,
for small $p$, than those of figure 1. Besides, it is important 
to check whether the main qualitative features of our results
are independent of the details of the infrared 
regularization scheme.
\paragraph{}

Fortunately, 
there is a 
consistent way of implementing the 
infrared cut-off within the ansatz (\ref{three}), which 
satisfies these criteria. We keep the limits of 
integration from $0$ to $\alpha$,and interpret the 
mass function $B/A$ in (\ref{one}) as a 
(covariant) infrared cut-off in the case of 
no dynamical mass generation. The expression for $A$ 
then reads
\begin{equation} 
A(p, \delta )=1 - \frac{\alpha}{\pi^2N}\frac{1}{p^3} \int 
_0^\alpha dk \frac{k}{k^2+ \delta ^2}I(p,k),
\label{onec}
\end{equation}
and the associated coupling $g_R(p,\delta)$ is defined by 
$g_R(p,\delta)=g_0/A(p,\delta)$.
 This way of introducing the infrared (IR) 
cut-off makes some contact with
the finite temperature case ~\cite{am,aklein}, where the 
plasmon mass was interpreted as an effective infrared cut-off.
The above similarity is, however, only indicative. Whether the 
situation described here carries over intact to the finite-
temperature regime is at present only an expectation.
These are issues that are left open for future investigations.
\paragraph{}

We may now repeat, for the ``$\delta$'' cut-off  case, the calculations
decribed above for the ``$\epsilon$'' cut-off. We first 
consider the use of (\ref{10b}) in (\ref{onec}), with 
the further restriction of keeping only the $\Theta (k-p)$ 
term, as in ~\cite{kondo}. This gives
\begin{equation}
g_R^{KN,\delta}=\frac{g_0}{1+\frac{g_0}{6} ln \frac{p^2+\delta^2}
{\alpha^2+\delta^2}},
\label{sixteen}
\end{equation}
which agrees with (\ref{grkn}) for $p > \epsilon$ as 
$\delta \rightarrow 0 $, and provides a smooth 
continuation to $p=0$. Next, we retain both $\Theta$ terms, as 
in (\ref{sdir}), leading to 
\begin{equation}
g_R^{(8),\delta}=g_0/A^{(8)}(p,\delta)
\label{seventeen}
\end{equation}
where (c.f. (\ref{a5}))
\begin{equation}
A^{(\ref{sdir})}(p,\delta)=1 - \frac{g_0}{9}
+ \frac{g_0}{3p^2}\delta ^2 -\frac{g_0}{3p^3}\delta ^3 
\arctan\left(\frac{p}{\delta}\right)+ \frac{g_0}{6}ln\left(
\frac{p^2 + \delta ^2}{\alpha ^2 + \delta ^2}\right).
\label{eighteen}
\end{equation}
Again we note that for fixed $\delta$ (\ref{eighteen})
goes smoothly to a constant value as $p \rightarrow 0$.
Finally, we calculate $g_R(p,\delta)$
numerically.
\paragraph{}
The results are shown in figure 2, which may be compared 
directly with figure 1. The first observation is that the 
appearance of the three curves is qualitatively very similar 
in the ``$\epsilon$'' and ``$\delta$'' cases, confirming that 
the detailed form of the infrared cut-off is not making a 
qualitative difference. Secondly,one sees that 
$g_R^{(8),\delta}$ runs parallel to $g_R^{KN,\delta}$ until 
$p$ is near $\delta$, and that both these approximations give 
significantly faster running than the exact $g_R(p,\delta)$.
Finally, we stress that there is a non-trivial frixed point 
at $p=0$, which is approached perfectly smoothly,in 
contrast to the unusual behaviour shown in figure 1. The result 
for $g_R(p,\delta)$ is therefore in full agreement with 
the expectations of I regarding a `walking technicolour'-like
behaviour of $g_R$ at intermediate scales of momenta.
\paragraph{}
To emphasize the insensitivity of these results
to the form of the infrared cut-off, we show again in figure 
3(a) the curves $g_R(p,\epsilon )$ of figure 1 and 
$g_R(p,\delta )$ of figure 2, for $ \epsilon =0.5$ and 
$\delta = 0.5$. Of course, there is no 
reason why such a comparison has to be made at the 
same numerical values of $\epsilon$ and $\delta$,
and we could have adjusted one of them to get 
closer matching of the results. But this would 
contain no extra physics: figure 3(a) already shows 
convincingly enough that the two crucial 
elements--the non-trivial fixed point at 
$p=0$ and the slow running--are independent of 
the form of the cut-off.
\paragraph{}
It interesting to note that the above features are 
characteristic of the presence of a {\em finite} 
infrared cut-off. Removal of $\delta$ via $\delta 
\rightarrow 0$ in a smooth manner does not seem to
be possible, as becomes clear from figure 3(b). From 
a physical point of view ~\cite{am}, where the 
infrared cut-off is conjectured to be 
associated with temperature in certain 
condensed-matter systems whose physics the above 
model is supposed to simulate, this would imply 
that the above non-trivial structure is an 
exclusive feature of the finite-temperature
field theory.
\paragraph{}
We add one final comment on the 
various approximations introduced above. Our
exact numerical results based on either
 (\ref{improvedeps})
or (\ref{onec}) show that the attractively 
simple formula (\ref{sdir}) (or the 
analogous one with the $\delta$ cut-off), which 
requires {\em both} of the approximations
(\ref{10a}) and (\ref{10b}), fails to capture the 
slow running at intermediate momenta. However,
approximation (\ref{10a}) alone does appear to 
be sound, and worth pursuing further. The expression 
for $A$ based on (\ref{10a}), with the $\delta$ cut-off,
is
\begin{equation}
A^{(6)}(p,\delta)=1-\frac{\alpha}{\pi^2N}
\frac{1}{p^3}\int_0^\alpha dk \frac{k}{k^2+\delta^2}
\{ f(p,k)+f(k,p) \}
\label{a55}
\end{equation}
where
\begin{equation}
f(p,k)={2k^3(4\alpha^3 p^3+4\alpha^2 p^4+\alpha^2 p^2 k^2
-3p^4 k^2+\alpha^2 k^4+3\alpha p k^4 + 3p^2 k^4) \over
(3 \alpha (\alpha p +p^2)^3)} \Theta(p-k),
\label{f}
\end{equation}
and we define the 
corresponding $g_R$ by $g_R^{(6)} = g_0/A^{(6)}(p,\delta)$.
The integrals in (\ref{a55}) can be performed analytically,
with the result:  
{\small
\begin{eqnarray}
&~&A^{(6)}(p,\delta) = 1- g_0
\bigl\{ {{{p^3}}\over {12\,{\alpha}\,{\delta^2}}} 
- {{{p^4}}\over {12\,{\alpha^2}\,{\delta^2}}} + 
   {{{p^2}\,\left( 2\,{\alpha^2} - {p^2} \right) }\over 
     {96\,{\alpha^2}\,\left( {\alpha^2} + {\delta^2} \right) }} - 
{{{p^2}\,\left( 2\,{\alpha^2} - {p^2} \right) }\over 
     {24\,\left( {\alpha^2} + {\delta^2} \right) 
\,{{\left( \alpha + p \right) }^2}}} + 
\nonumber \\
&~&      {{\left( p-{\alpha} \right) \,\left( 4\,{\alpha^6} + 4\,{\alpha^4}\,
{\delta^2} - 
	 2\,{\alpha^4}\,{p^2} - 6\,{\alpha^2}\,{\delta^2}\,{p^2} 
- {\alpha^2}\,{p^4} + 
	 {\delta^2}\,{p^4} \right) }\over 
     {24\,{\alpha}\,{{\left( {\alpha^2} 
+ {\delta^2} \right) }^2}\,\left({\alpha} 
+ p \right)}} + \nonumber \\
&~& {{140\,{\alpha^2}\,{\delta^4}\,{p^2} 
  -105\,{\alpha^2}\,{\delta^6} - 315\,{\alpha}\,{\delta^6}\,p 
       - 315\,{\delta^6}\,{p^2} - 
       420\,{\alpha^3}\,{\delta^2}\,{p^3} 
+ 105\,{\alpha}\,{\delta^4}\,{p^3} \over 
     {1260\,{p^5}\,{{\left( \alpha + p \right) }^3}}}} + 
\nonumber \\  
&~& {{140\,{\alpha^3}\,{p^5} - 476\,{\alpha^2}\,{\delta^2}\,{p^4} 
       - 210\,{\delta^4}\,{p^4} - 
       63\,{\alpha}\,{\delta^2}\,{p^5} + 
   176\,{\alpha^2}\,{p^6} + 42\,{\delta^2}\,{p^6} + 
       45\,{\alpha}\,{p^7} - 18\,{p^8}}\over 
     {1260\,{p^5}\,{{\left( \alpha + p \right) }^3}}} 
+ \nonumber \\  
&~& {{{\delta^3}\,\left( {\alpha^2}\,{\delta^4} 
+ 3\,{\alpha}\,{\delta^4}\,p - {\alpha^2}\,{\delta^2}\,{p^2} + 
	 3\,{\delta^4}\,{p^2} + 4\,{\alpha^3}\,{p^3} + 4\,{\alpha^2}\,{p^4} + 
	 3\,{\delta^2}\,{p^4} \right) } 
\over 
     {12 \, {p^6} \, {{\left( \alpha + p \right) }^3}}} 
\arctan \left({p\over \delta} \right)  + \nonumber \\
&~& {{ 8 {\alpha^5} {\delta^4} 
+ 8 {\alpha^3} {\delta^6} + {\alpha^5} {\delta^2} {p^2} - 
	 9 {\alpha} {\delta^6} {p^2} 
- {\alpha^5}\,{p^4} - 3 {\alpha^3} {\delta^2} {p^4} - 
	 6 {\alpha} {\delta^4} {p^4} } \over 
     {12\,{\delta^3}\,{{\left( {\alpha^2} + {\delta^2} \right) }^3}}} 
       \left( \arctan \left({\alpha\over \delta} \right) 
- \arctan \left({p\over \delta} \right) \right) 
+ \nonumber \\
&~& {{ 4\,{\alpha^2}\,{\delta^4} - 4\,{\alpha^6}  
+ 3\,{\alpha^4}\,{p^2} + 
	 8\,{\alpha^2}\,{\delta^2}\,{p^2} 
- 3\,{\delta^4}\,{p^2} + {\alpha^2}\,{p^4} - 
	 3\,{\delta^2}\,{p^4} }
     \over {24\,{{\left( {\alpha^2} + {\delta^2} \right) }^3}}}
       ln \left({{4\,{\alpha^2}\,\left( {\delta^2} + {p^2} \right) }\over 
     {\left( {\alpha^2} + {\delta^2} \right) 
\,{{\left( \alpha + p \right)}^2}}} \right) 
\bigr\}
\label{analyt}
\end{eqnarray}
}
The 
interesting point is that $g_R^{(6)}(p,\delta)$ turns out to 
be almost indistinguishable from the exact $g_R(p,\delta)$,
showing that (\ref{10a}) is actually very reliable.
This is shown in figure 4. 
The availability of 
this very accurate analytical expression for $g_R$ might be 
useful. For example, taking the derivative of (\ref{analyt})
with respect to 
$p$ it is straightforward to show that the non-trivial
infrared fixed point does lie at $p=0$.
 It is probably of interest to the 
detailed reader to  note that 
the more crude approximation consisting of keeping 
only the $\Theta (k-p)$ terms in (\ref{10a}),
which is an improved version of 
what was done in \cite{kondo}, 
exhibits a much better agreement with the exact 
answer of fig. 2 and (\ref{analyt})
than that based on the approximation 
(\ref{grkn}). This shows that the inadequacy 
of the approximation of ref. \cite{kondo} lies mainly 
in the steps leading to (\ref{10b}), rather than in the 
neglect of modes with momentum below $p$. 
\paragraph{}
To conclude, in this short note we have 
solved the SD equations (\ref{one}), with the ansatz $n=1$ in 
(\ref{three}),
 for the wave-function
renormalization in the normal phase of the model, where dynamical
mass is ignored. The solution has been exact, in the sense that 
it is valid for a wide region of momenta $0 < p < \alpha$.
The solution indicated that the increase of the running coupling
is cut-off in the infrared, in the form of a non-trivial fixed 
point, and that there is also a significant slowing down 
of the running of the coupling at intermediate scales,
as compared to the case of ref. \cite{kondo} or of 
Eq. (\ref{gr5}).
This slowing down, or `walking-technicolour-like' behaviour,
has been argued~\cite{am} to be responsible for the 
smallness of the coherence length of the 
dynamical-mass-generation phase of the theory.
On the other hand, both features, the non-trivial 
infrared structure and the walking-coupling behaviour
at intermediate scales, have been  argued in I 
to be responsible for deviations from fermi-liquid 
behaviour, which might have important 
physical consequences,  in case the model  
simulates correctly  the physics of the novel high-temperature
superconductors. We have also argued that the above 
non-trivial low-energy structure is a consequence of a 
{\it finite} infrared cut-off. Thus, from the physical
point of view, adopted in I, of associating the infrared cut-off
with temperature effects, this would imply 
that the above structure is an exclusive feature 
of the finite-temperature field theory. 

\vglue 0.6cm
\leftline{\Large {\bf Acknowledgements}}
\vglue 0.4cm

We would like to acknowledge discussions with L. Del Debbio, 
S. Hands, and K. Kondo. 
G.A.-C. acknowledges financial support 
from the European Union  
under contract \#ERBCHBGCT940685.
M.K.-K. thanks the Royal Society of London and the Mexican Academy
of Scientific Research (AIC), under a joint collaboration programme,
for supporting financially a visit to the University of Oxford,
during which the present work was carried out.
D.McN. wishes to thank PPARC for a research studentship.

\newpage
\paragraph{}
\noindent {\Large {\bf Figure Captions}}
\paragraph{}
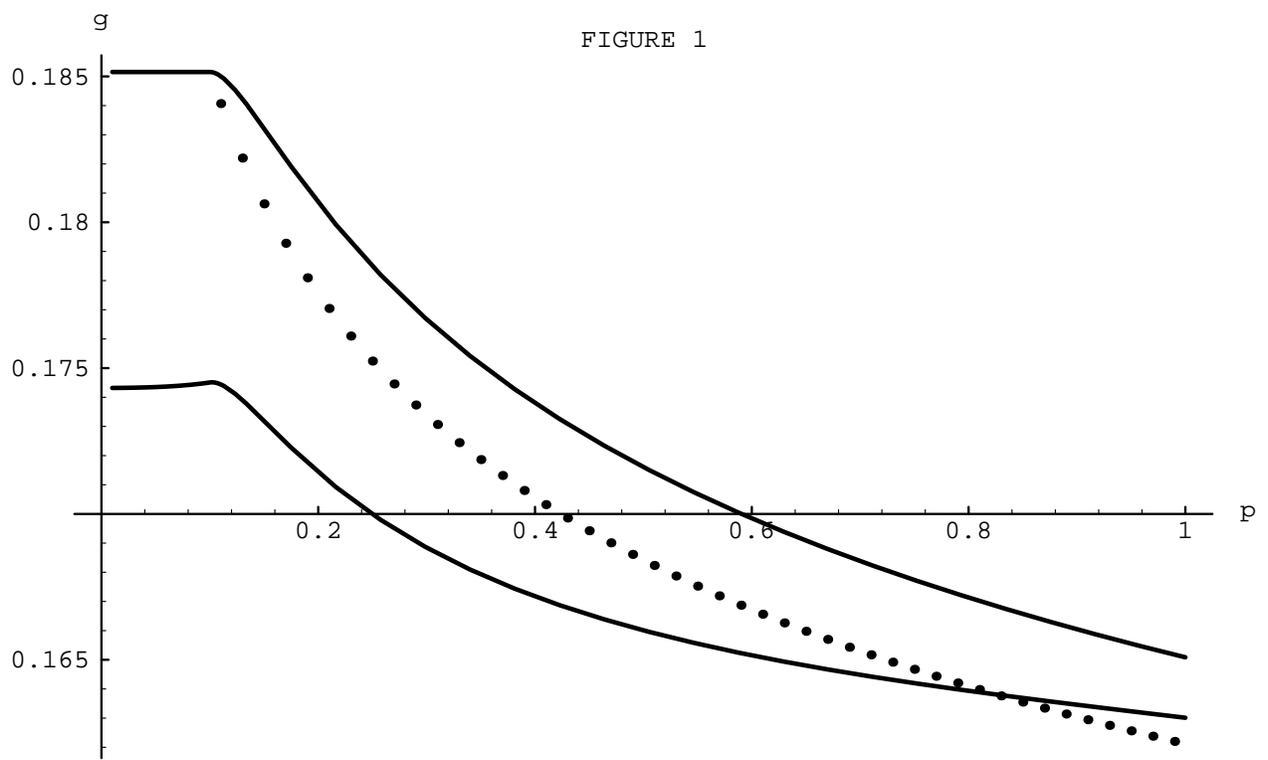
\noindent {\bf Figure 1}: Running coupling constant of large-N resummed 
$QED_3$  in the normal phase 
as 
a function of the external momenta $p$, in the 
presence of a momentum-space infrared cut-off $\epsilon$. 
In the figure  
$\epsilon=0.1$, $\alpha =1$,
and $N=5$;
the dotted curve shows $g_R^{KN}$ of (\ref{grkn});
the top continuous curve shows $g_R^{(8)}$ of (\ref{gr5})
and (\ref{a5});
the bottom continuous curve shows the exact (numerical)
result, $g_R(p,\epsilon )$.
\paragraph{}
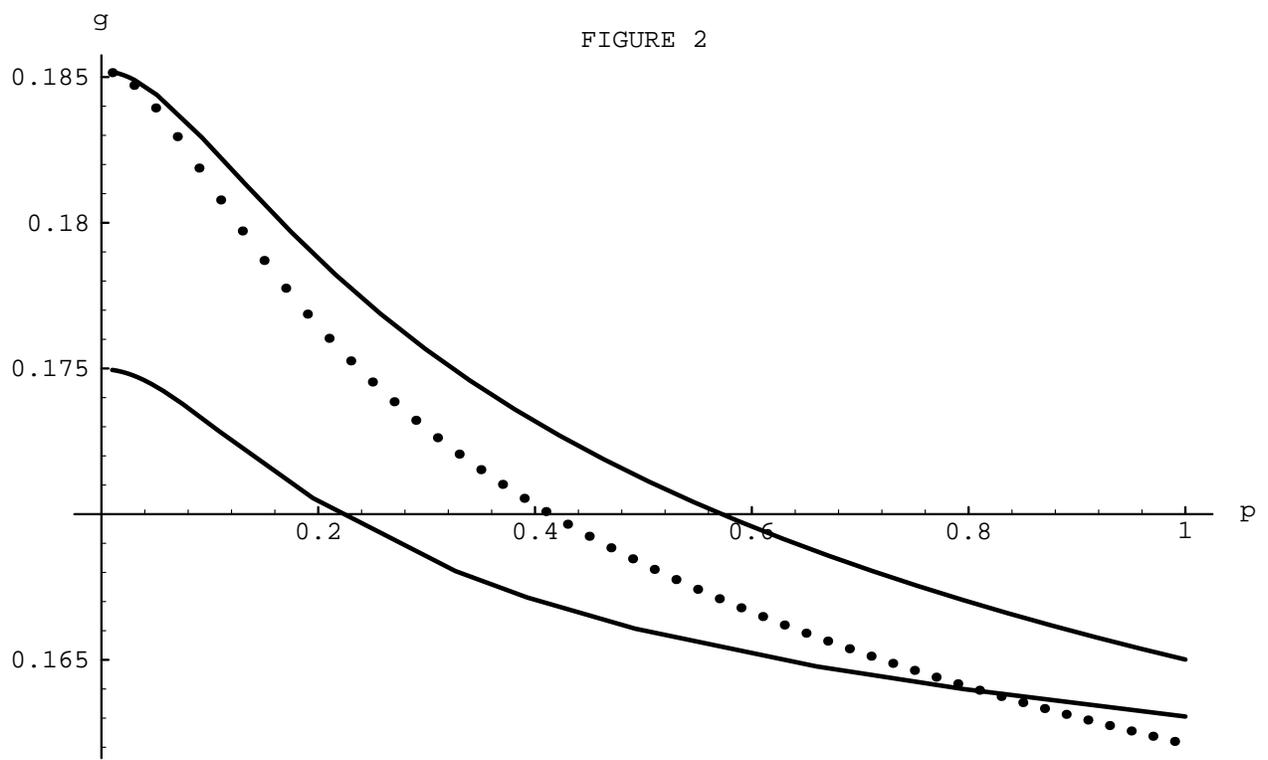
\noindent {\bf Figure 2}: As in figure 1, but now employing 
a covariant infrared-mass 
cut-off $\delta =0.1$. The dotted curve is 
$g_R^{KN,\delta}$ of (\ref{sixteen}); the top 
continuous curve is $g_R^{(8),\delta}$ of (\ref{seventeen});
the bottom continuous curve is $g_R(p,\delta)$.

\paragraph{}
\noindent {\bf Figure 3}: A comparative study of the running of the 
$QED_3$  coupling in the cases with momentum-space ($\epsilon$)
and convariant-mass ($\delta $) infrared cut-offs 
for $\alpha = 1$, $N=5$. 
In 
(a) the dotted and continuous 
curves show $g_R(p,\epsilon =0.1)$ and
$g_R(p,\delta =0.1)$ respectively. In (b) 
 is shown the  
value of the coupling $g_R(p=x,x)$
at momenta of order of the generic
infrared cut-off $x$, as a function of the cut-off
(the dotted curve corresponds to the $\epsilon$ cut-off, whilst 
the continuous one corresponds to the $\delta$ cut-off).  
Clearly the cut-off cannot be removed smoothly in either case. 
\paragraph{}
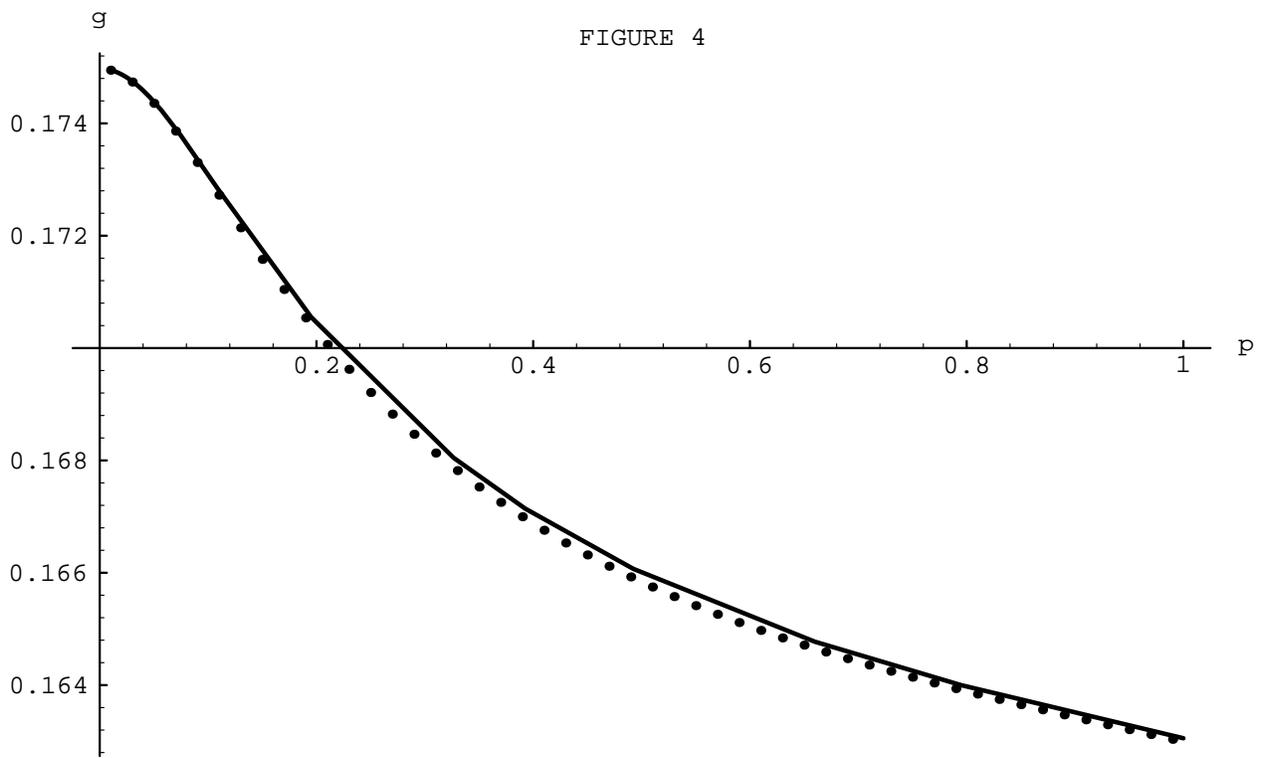
\noindent {\bf Figure 4}: Test of the validity of the 
analytic 
result (\ref{analyt}), based on the 
approximation (\ref{10a}),
in the covariant mass-cut-off $\delta $ case. 
The running of the coupling based on (\ref{analyt}) (dotted curve)
is compared with the exact numerical result (continuous 
curve), for $\alpha =1$, $\delta = 0.1 $ and $N=5$. 
The agreement 
is very good throughout the entire range of 
external momenta $0< p < 1$.

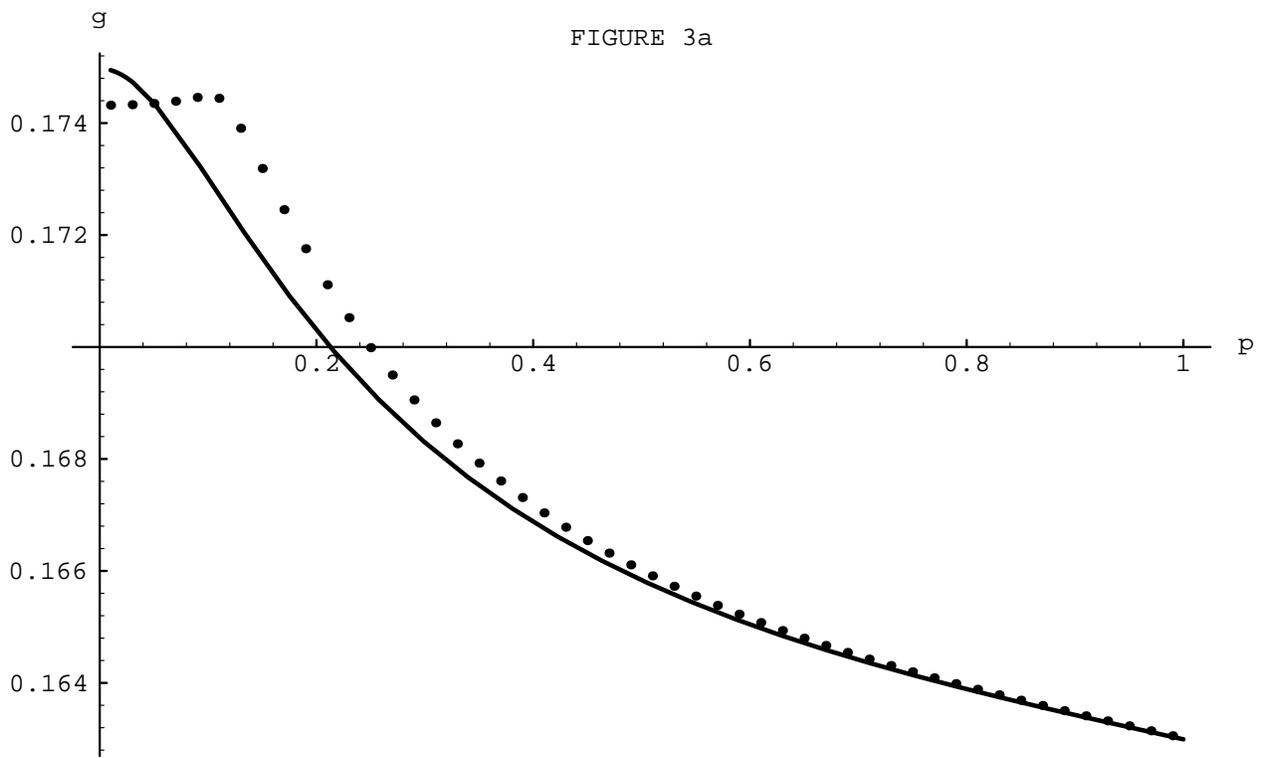
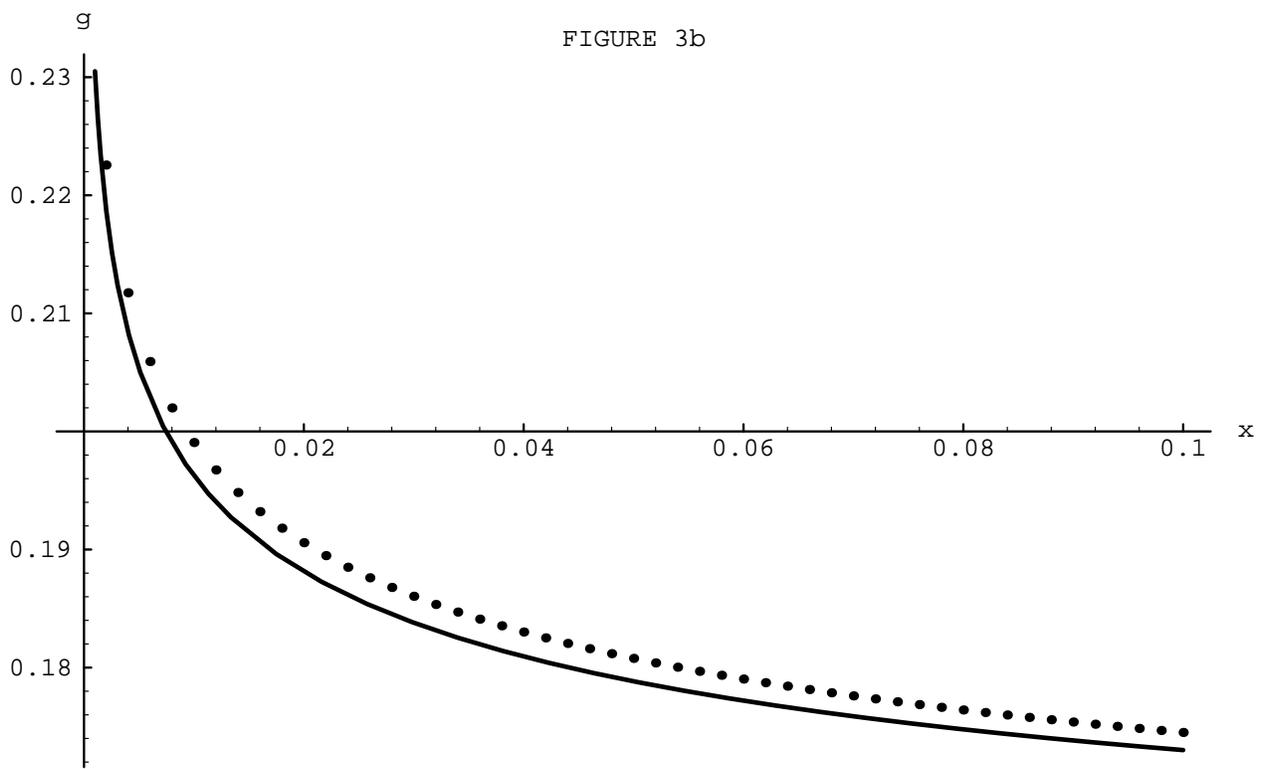
\end{document}